\documentstyle[preprint,aps]{revtex}
\tightenlines
\begin{document}
\title{An Effective-Medium Tight-Binding Model for Silicon}

\author{K. Stokbro, N. Chetty$^1$, K. W. Jacobsen, and J. K. N\o
rskov}
\address{
Center for Atomic-scale Materials Physics and Physics Department, \\
Technical University of Denmark, DK 2800 Lyngby, Denmark \\
$^1$ Department of Physics, Brookhaven National Laboratory \\
 Upton, NY\ \ 11973, USA}

\maketitle

\begin{abstract}
A new method for calculating the total energy of Si systems is
presented. The method is based on the effective-medium theory concept
of a reference system. Instead of calculating the energy of an atom in
the system of interest a reference system is introduced where the
local surroundings are similar. The energy of the reference system
 can be calculated
selfconsistently once and for all while the energy difference to the
reference system can be obtained
approximately. We propose to calculate it using the tight-binding LMTO
scheme with the Atomic-Sphere Approximation(ASA) for the potential, and by
using the ASA with charge-conserving spheres we are able to treat open system
without introducing empty spheres.
All steps in the
calculational method is {\em ab initio} in the sense that all
quantities entering are calculated from first principles without any
fitting to experiment. A complete and detailed description of the
method is given together with test calculations of the energies of
phonons, elastic constants, different structures, surfaces and surface
reconstructions. We compare the results to calculations using an
empirical tight-binding scheme.
\end{abstract}

\section{INTRODUCTION}

Two parallel developments have  changed our possibilities of
understanding and predicting the energetics and dynamics of
condensed-matter systems over
the past few years. One is the development of {\em ab initio}
total-energy methods\cite{carpar} based on  density-functional
theory\cite{hohenberg}, and the
local-density approximation\cite{ldarev}. It is now possible to calculate total
energies
and equilibrium configurations of systems with up to a few hundreds of
atoms\cite{si(7x7)}, and in other cases it has been possible to study
the molecular dynamics directly\cite{carpar,cl2diss,blochl}. During this
time, there has been a
parallel development of approximate and semi-empirical total-energy
methods\cite{emt,stott,daw,sinclair,glue,jacobsen,depristo}   which has made it
possible to treat the dynamics and thermal
properties of systems with many  thousand atoms with a reasonable
accuracy\cite{cucluster,uzi}. With such methods it is now possible to treat
problems in
materials physics where extended defects or long range disorder are
crucial for the properties that are under study.

The hope is to  develop methods with the accuracy and
robustness of the {\em ab initio} methods which can handle the
larger systems on a reasonable time scale. One problem with the {\em ab
initio} methods is the fact that the computer time scales as the cube
of the  number of atoms. New developments of methods where the computer time
scales linearly with the size of the system have all focussed on
localized basis sets\cite{orderN1,orderN2}. One important  problem here is to
construct the Hamiltonian. A natural choice would be to use the LMTO
method in the tight-binding formulation\cite{tblmto}, but at the
present time these methods  usually rely on semi-empirical
tight-binding models constructed to fit a set of properties.

In the present paper we introduce a very efficient and accurate method
for calculating total energies for Si based on density-functional
theory. The method is approximate, but computationally very efficient
(2 orders of magnitude faster than conventional {\em ab initio}
methods). A set of well defined approximations are made in the total
energy expression
where we rely on the variational nature of the generalized total energy
functional to compute energies reliably. Since the approximations are very
systematic, we are able to test  the validity of the assumptions at each stage,
and therefore in a controlled way develop  a hierarchy of models with various
levels
of approximations. All input terms are calculated theoretically, i.e. no
fitting to experiment. We have earlier demonstrated the versatility of
the first levels of approximations in its application to metallic
Al\cite{abinitio}.

 The method utilizes the
effective-medium theory concept of an effective medium or reference
system. Instead of calculating the total energy of a system of
interacting Si atoms directly, we associate with each atom in the
system a reference system where the energy of the atom can be
calculated easily and where the surroundings of the atom in question
are sufficiently similar
to those of the system of interest that the energy difference can be
calculated approximately. Our main approximations are to use
a transferable input charge density\cite{chetty} and a transferable effective
potential\cite{optpot}. From these approximations the energy
difference can be calculated with a density-dependent pair potential and
 an LMTO tight-binding  Hamiltonian, with all the
in-going  parameters determined from the transferable charge density
and effective potential.

We have made extensive comparisons of the results of this new method with
selfconsistent calculations of elastic properties, phonons, surfaces, different
crystallographic phases, surface reconstructions and adatoms on surfaces.
In all cases the quality of the results is good, and we show that in the cases
where the empirical tight-binding method is known to fail, the new method
works well.

The main result of the present paper is the scheme to calculate total
energies for Si systems, which is summarized in the first part of section IV.
The basis of the method is
the Harris functional and the effective-medium
theory concept of a reference system. These aspects of the paper are
discussed in section II and the first part of section III.
In the second part of section III we construct the LMTO tight-binding
model from which the one-electron energy is calculated.
The tests
of the new method are given in section IV, which also includes a
discussion of the relation of the present method to the empirical
tight-binding method.

\section{THE FIRST LEVEL OF APPROXIMATION: THE OPTIMIZED DENSITY AND
HARRIS FUNCTIONAL}
\subsection{General remarks}

 The so-called Harris functional \cite{harris} is  a good starting point for
 investigating the theoretical foundations of the tight-binding
method\cite{foulkes,sankey}
due to  the non-selfconsistent nature of the functional
and the fact that it only depends on the
input charge density. If the input density is good enough
\cite{finnis},
then this will give a considerable
savings in computer time since only a single iteration of the Kohn-Sham
equations is needed. In this work, we invoke these properties of the
Harris functional to develop a model potential for Si.

\subsection{Constructing atomic-like optimized densities}
We have discussed previously \cite{chetty,robertson2} the systematic
decomposition of a selfconsistent total charge density into atomic-like
optimized densities
\begin{equation}
n({\bf r})\, =\, \sum_{{\bf R}_{\mu}}\, \Delta n_{op}({\bf
r}-{\bf R}_{\mu})\, .
\label{eq:opt}
\end{equation}
Here, we use norm-conserving nonlocal pseudopotentials
\cite{shirley,vanderbilt} within a plane-wave basis \cite{ihm} at an energy
cutoff of 12 Ry to compute the selfconsistent
charge densities of bulk Si and the ideal (111) surface from which we extract
the reciprocal-space components of the optimized density $\Delta n_{op}(k)$
(see Appendix I). Reverting to real-space, the optimized density shown in
Fig. 1 has the well known features of a contraction in the outer core region
and
a sharper attenuation of the tail with some barely discernible Friedel
oscillations \cite{puska}.
As noted before \cite{chetty}, embedding an atom in a homogeneous electron
gas at typical metallic densities produces very similar features, and this
is of interest here because renormalizing the atom in this manner is the basis
of the effective-medium theory of Jacobsen, Puska and N\o rskov
\cite{jacobsen}.

We have extracted reciprocal-space components of the optimized density for the
other two principal surface orientations, viz. (100) and (110), and we find
that to a good approximation, the components fall on the
same universal curve. We will now make the assumption that $\Delta n_{op}(r)$
is indeed universal and transferable, and in the next subsection we will
test this ansatz by computing the Harris functional for various
different test situations and comparing with the selfconsistent results.

\subsection{Results}

In order to test the accuracy of the approximations at each step in the
 potential construction, we will use a data base of test systems. This
 data base covers silicon in different crystallographic structures at the
 equilibrium volume, phonons, elastic constants, and surfaces. For the
 surfaces, we have used a supercell geometry with 12 atoms, and for all
 calculations we ensure an adequate sampling \cite{monkhorst} of the Brillouin
 zone.

 In Table~1, we compare the Harris with the selfconsistent results. We also
 include the effective-medium tight-binding (EMTB) results and those of
the empirical tight-binding model of
Ref.~\cite{pettb}, which we
 discuss in the next section. We note that using the Harris functional
with  our choice of input-charge
 density, constructed from spherically-symmetric atomic-like densities
(i.e. without bond-charges),  there is excellent corroboration with
 the selfconsistent results. Other studies\cite{read,polatoglou} using the
 free atom density as an input into the Harris functional have reproduced the
 bulk selfconsistent results with a similar degree of accuracy.

Our first new result is that the total energies of the
ideal surfaces of this semiconducting material are accurately derived
 from a non-selfconsistent calculation. It is interesting to note that this is
 so despite the fact that the exact position of the surface states are known to
 be sensitive to the degree of selfconsistency. The reason is simply that the
 total energy is stationary in the density while the position of the surface
 states are not. We therefore emphasize that the derived potential should only
 be used for predicting physical quantities which are stationary in the
density.

\section{THE SECOND LEVEL OF APPROXIMATION: THE EFFECTIVE-MEDIUM
TIGHT-BINDING MODEL}
\subsection{general remarks}
We will use the effective-medium construction as a basis for making further
approximations. The effective-medium idea is to
first  calculate the total energy for a series of reference systems, and then
 for a given system  relate each atom to a reference system, and only calculate
the energy difference between the two systems. If the reference
system is chosen wisely, the energy difference will be
small, and can therefore be calculated approximately. To find the appropriate
reference system, we will introduce the concept of a neutral sphere,
defined to be a sphere around an atom for which the electron density exactly
compensates the positive atomic charge.
As reference system we will use silicon in the diamond structure
with a lattice constant such that the
neutral-sphere radius  is the same in the reference system
as for the atom in the original system.

In calculating   this energy difference
 we will utilize that
the Hohenberg-Kohn density functional can be generalized to a
functional $E[n,v]$ which depends on both the density $n$ and the
potential $v$\cite{jacobsen,foulkes} and which is stationary with
respect to independent variations of each variable. The
general functional can be written as
\begin{equation}
E[n,v]= \sum_{\alpha} \epsilon_\alpha[v] - \int n({\bf r}) v({\bf r})  d{\bf r}
+ E_{el}[n] + E_{xc}[n],
\end{equation}
where $\epsilon_\alpha[v]$ denotes the eigenvalues generated by
solving the Kohn-Sham equation with the potential
$v$, and where $E_{el}[n]$ and $E_{xc}[n]$ is the electrostatic and
exchange-correlation energy, respectively. If the potential is restricted
to be a functional of the density the Hohenberg-Kohn functional or the
Harris functional appear as special cases\cite{jacobsen}.

We have already used the stationary property of the density
by calculating the total energy
using a superposition of atom-based optimized densities. We will show that
from this approximation  the electrostatic and exchange-correlation
energy can be transformed into a density-dependent pair-potential sum by
linearizing the
exchange-correlation functional.

In order to get a simple scheme for calculating the kinetic energy
we will use the Atomic-Sphere Approximation (ASA). We thereby exploit the
stationary property of the  the kinetic energy functional with respect to
 variations in the potential by substituting the full potential with
the spherically-averaged potential within each neutral
sphere. Furthermore, we have recently shown\cite{optpot} that
the spherically-averaged potential
 of the reference system is very similar to that of the real system,
and since the kinetic energy functional is stationary in the potential
 it is a good approximation
to substitute the spherically-averaged potential within each sphere
with  the potential of the
 reference system. This last approximation transforms the potential-energy
 contribution to  the kinetic-energy difference into
a sum of density-dependent pair potentials.

The remaining term is the one-electron energy which we calculate using
an LMTO tight-binding Hamiltonian.
 Since the only potential appearing in the problem is now that of the reference
 system, we can precalculate the potential parameters and, by constructing an
 interpolation formula for the structure constants, we obtain a simple
 density-dependent two-center tight-binding Hamiltonian.

 We have now given a brief description of the main approximations used in the
 construction of the Effective-Medium Tight-Binding (EMTB) model. In the
 following, we give a more detailed account of the construction.

\subsection{The diamond reference system and the neutral-sphere radius}

 In the original effective-medium theory of Ref.~\cite{emt}, each atom is
 viewed as embedded in the electron density from the neighboring atoms and,
 when averaged, this density provides an effective medium in the form of a
homogeneous electron gas.
The role of the homogeneous electron gas
is to provide a reference system in which the atoms have similar
chemical surroundings as in the original system.
In order to obtain the total energy, corrections
due to the non smoothness of the charge density have to be included;
however these
corrections are small and can be calculated approximately.
 We will use the effective-medium construction and calculate the total
energy $E$ as
\begin{equation}
E = \sum_i e^{ref}(s_i) +[E- \sum_i e^{ref}(s_i) ],
\label{eq:emt}
\end{equation}
where we use the neutral-sphere radius
 $s_i$ of each atom to define the reference system and $e^{ref}(s)$
 is the energy of the reference system with
neutral-sphere radius $s$. Note that since
the electron density is constructed using Eq.~(\ref{eq:opt}),
the choice of identical neutral spheres
in the two systems is equivalent to the choice of identical embedding density
used in Ref.~\cite{emt}.

For silicon in the pseudo-potential scheme the neutral sphere
contains 4 electrons, and from
Eq.~(\ref{eq:opt}) we obtain the equation
\begin{equation}
\label{eq:neuts}
4  =  \sum_{j} \Gamma(d_{ij},s_i) \;  ,
\end{equation}
where $\Gamma$ is the electron-density contribution from atom $j$
to a sphere at site $i$ of radius $s_i$,
\begin{equation}
\Gamma(d,s)  =  \int_{s} d{\bf r}
\Delta n_{op}(|{\bf r} - {\bf d}|).
\label{eq:gamma}
\end{equation}
 We solve Eq.~(\ref{eq:neuts}) iteratively for the neutral-sphere
radius, using  a cutoff distance
 of $r_c = 11.67 a_0$ to terminate the sum (we thereby
include five neighbor shells in the equilibrium  diamond lattice). For this
lattice, the nearest-neighbor distance is $d_0=4.40 a_0$ and we obtain a
neutral-sphere radius of $s_0=2.72 a_0$.
In Fig.~\ref{fig:dsss} we show
the neutral-sphere radius of the diamond structure as a function of the
nearest-neighbor distance. For comparison, we also show the Wigner-Seitz
radius of the diamond lattice and the $2^{\frac{1}{3}}$ smaller
Wigner-Seitz radius of a diamond lattice embedded in a bcc lattice
with empty spheres.
We see that  in the diamond structure the neutral-sphere radius is
substantially
smaller than the Wigner-Seitz radius. This difference is due
to the large regions in the diamond lattice which contain almost no charge
and therefore are not included in the neutral sphere.
In a more close packed system like the fcc system the neutral-sphere radius
is more or less equal to the Wigner-Seitz (WS) radius.

As reference system we will use the diamond structure, and
the energy correction therefore vanishes for
a diamond lattice or an isolated atom, since the latter can be regarded
as a diamond lattice with an  infinite lattice constant.
So by construction the EMTB will give the
correct cohesive energy, bulk modulus, and equilibrium lattice constant for
the diamond structure.

In order to calculate the energy difference to the reference system,
 we divide the total
energy into two terms;  The kinetic energy ($T$) and the sum of the
electrostatic and exchange-correlation energy  ($G$)
\begin{eqnarray}
 E & = & \sum_i e^{ref}(s_i) + [T-\sum_i t^{ref}(s_i)] +
[G-\sum_i g^{ref}(s_i)], \\
 E & = & \sum_i e^{ref}(s) + \Delta T + \Delta G. \nonumber
\end{eqnarray}
So far we have just rearranged the terms in the total energy.
The first approximation we make is to calculate the
kinetic energy in the atomic-sphere approximation (ASA).
The quality of this approximation depends on how well the full potential
can be approximated by a superposition of
 spherically-averaged potentials within atomic spheres.
Traditionally, the ASA is made
using  space-filling spheres on the ground that with this choice integrals
over space may be mapped into integrals over the spheres --
 the
integration of a constant function will therefore be correct in the ASA.
 However,  since the density enters all integrals, we will
use neutral spheres, making the spheres charge conserving instead of volume
conserving --  the integration of a
 constant function times the density is correct.
This choice seems more physical since
we thereby obtain that both the ASA density and the full density
contains the correct number of electrons, which is not fulfilled when the ASA
is used with volume conserving spheres.
In Fig.~\ref{fig:sipot} we show the full selfconsistent
potential of silicon in a (110) diamond plane, together with the
neutral-sphere radius (solid circle)
and Wigner-Seitz radius (dotted circle). It is clear from the figure that the
overlap region
of the WS spheres penetrates far into the spherically symmetric parts
of the potential,
which would lead to a poor approximation of the full potential.
The overlap region
of the neutral-spheres, on the other hand, very accurately sample
 only the asymmetric part of the potential and since  the potentials are
superimposed the approximation is enhanced in this region. We will use the ASA
with
neutral spheres, and because we thereby obtain an accurate approximation for
the full potential we may avoid the traditional usage of empty spheres in
ASA calculations for open structures.

We have recently shown that with our choice of reference system the
ASA potential of a general system is almost identical to the potential
of the reference system\cite{optpot}, so that  we may at each site substitute
the
potential within the atomic sphere with the reference potential. Furthermore,
due to the variational properties of the energy functional, this only
leads to errors in the total energy of second order in the potential
difference $(v-v^{ref})$.
 We then  have for the kinetic-energy difference
\begin{eqnarray}
\label{eq:tapp}
\Delta T & \approx & \sum_{\alpha \in occ} \epsilon_{\alpha}[\bar{v}^{ref}] -
\sum_i e_{1el}(s_i)  - \sum_i \int_{s_i} \bar{v}^{ref}(r,s_i)
(n({\bf r})-n^{ref}({\bf r})) d{\bf r}, \\
& \equiv & \Delta E_{1el} + \Delta V, \nonumber
\end{eqnarray}
where $\bar{v}^{ref}$ is the spherically-averaged potential of the
reference system, and
$e_{1el}(s)$ is the one-electron energy of the reference system
with neutral-sphere radius $s$.

Let us now summarize the total binding-energy expression in the form used
in the Effective-Medium theory of
Ref.~\cite{jacobsen}. The total binding energy is given by
\begin{eqnarray}
\label{eq:etot}
E  & = & E_c + \Delta E_{as} + \Delta  E_{1el}, \\
 & = & \sum_i e^{ref}(s_i) + [ \Delta G + \Delta V ] + \Delta E_{1el},
\nonumber
\end{eqnarray}
where $E_c,\; \Delta E_{as}$, and $\Delta E_{1el}$ are called the cohesive
function, atomic-sphere
correction, and one-electron correction, respectively.

The first term, the cohesive function,
is the energy of the reference system, which we parametrize
using the interpolation formula\cite{unibind}
\begin{equation}
E_c(s) = E_0 ( 1 + x) e^{-x},\; \; x = \lambda (s -
s_0).
\label{eq:unibind}
\end{equation}
In this equation, $E_0 = -5.83$, is the cohesive energy of the equilibrium
diamond
lattice and the parameter,
$\lambda = 2.047$, is determined by the bulk-modulus of the diamond lattice.

The second term, the atomic-sphere correction, can be viewed as a
correction to an ASA calculation
of the electrostatic and exchange-correlation energy. To see this we use
the definition of the effective potential
$\bar{v}^{ref} = \frac{\partial{\bar{g}^{ref}}}{\partial{n}}$, where
$\bar{g}^{ref}$ is the ASA electrostatic and exchange-correlation
energy of  the reference system. We can now identify the
$-\Delta V$ term of Eq.~(\ref{eq:tapp}) as the first term in a Taylor
expansion of the difference in the
ASA electrostatic and exchange-correlation energy between the system
and the reference systems ($\Delta \bar{G}$), and we therefore have
$\Delta E_{as} = \Delta G -{\Delta \bar{G}} + O([n-n^{ref}]^2)$. We
see that the atomic-sphere correction is the first order correction to
a calculation where not only the kinetic energy, but also the exchange
and correlation energy have been calculated within the ASA, i.e. the
ASA has been used for both the potential and the density.
In the next section we will show that
the atomic-sphere correction can be calculated by a
density-dependent pair potential.

The third term,
the one-electron correction, is the energy correction due to the difference in
band structure between the system and the reference system, and in
section~\ref{sec:e1el} we will
calculate this term using an
LMTO tight-binding model.

In Table~2 we show the EMTB terms for the  test systems of Table~1.
The cohesive energy and atomic-sphere correction were extracted from a Harris
functional calculation and  the one-electron correction then obtained by
subtracting these terms from the total energy. With this data base we
may not only check the potential by calculating the total energy, but we
may  test the accuracy of each term in the energy separately.
For now we will just note
that the contribution from the cohesive function to the energy
 of the equilibrium structures is very small, indicating that it is
the minimum of this function that determines the equilibrium
volumes. Since the cohesive function depends only on the
neutral-sphere radius, this implies
 that all the  equilibrium structures have almost the same neutral-sphere
radius, even though the equilibrium  volumes are very different.

\subsection{Calculating the atomic-sphere correction with a
density-dependent pair potential}
We will now decompose the atomic-sphere correction into
density-dependent pairwise interactions. The interactions depend on
the local density through the neutral-sphere radius s
\begin{equation}
E_{as}  =   \sum_{i,j} V(d_{ij}, s_i) ,
\label{eq:easv}
\end{equation}
where the pair potential is composed of three parts
\begin{equation}
V(d,s)  =  V_v(d,s) + V_{el}(d) + V_{xc}(d,s),
\label{eq:vparpot}
\end{equation}
which originates from the $\Delta V$ term,
the electrostatic energy, and the exchange-correlation energy, respectively.
{}From Eq.~(\ref{eq:opt}) and Eq.~(\ref{eq:tapp}) we see that
 Eq.~(\ref{eq:easv}) is exact for
the $\Delta V$ term and the electrostatic energy,
with the pair potentials given by
\begin{eqnarray}
V_v(d,s) & = & \int_{s} \bar{v}^{ref}(r,s) \;
 \Delta n_{op}(|{\bf r} - {\bf d}|) d{\bf r}, \\
V_{el}(d) & = & \int  [ v_{l}(r) + \frac{1}{2}
\int\frac{\Delta n_{op}(r')}{|{\bf r} - {\bf r'}|} d{\bf r'}] \;
 \Delta n_{op}(|{\bf r} - {\bf d}|) d{\bf r}.
\end{eqnarray}
We only include the local part of the pseudo potential $v_l$, since
the nonlocal part is canceled between the electrostatic energy and
the potential part of the kinetic energy.

To calculate the exchange-correlation energy we use the local-density
functional of Ref.~\cite{lda}, and from the decomposition of the
density we have
\begin{equation}
E_{xc} = \sum_i \int \Delta n_{op}(|{\bf r}-{\bf R}_i|) \, {\cal
E}_{xc}(n({\bf r})) \, d{\bf r}.
\end{equation}
 In order to approximate  this term with
a pair-potential sum we will have to divide the exchange-correlation
functional into contributions from each atom, which may be obtained by
using a linear expansion for the exchange-correlation functional.
We have chosen to linearize the exchange-correlation functional around
the spherically-averaged density of the reference
system, $\bar{n}(r,s)$,  and thereby  obtain
the pair potential
\begin{equation}
V_{xc}(d,s) = \int \Delta n_{op}(r) \frac{\partial {\cal
E}_{xc}}{\partial n} (\bar{n}(r,s))
\Delta n_{op}(|{\bf r} - {\bf d} |) d {\bf r}.
\end{equation}
 In  Table~\ref{tab:efitxc}
we compare the exchange-correlation part of the atomic-sphere correction
calculated using this pair potential to an exact evaluation of the
exchange-correlation
integral.
We see that the correction generally is underestimated by 20 percent.
This suggests that a better approximation might be obtained if
the exchange-correlation functional is
linearized at a lower density than the mean density, possibly because it
varies more rapidly in the low density regime.
We have not addressed this problem further, being satisfied
with the fact that  when a factor
is allowed for rescaling the pair potential we obtain a good description of the
exchange-correlation energy, as shown in the third
column of Table~\ref{tab:efitxc}.

We conclude this section by showing in Fig.~\ref{fig:vd} the distance
dependence
of the pair potentials, and in Fig.~\ref{fig:vs} the dependence upon the
neutral-sphere radius. We see that the distance dependence of the sum is nearly
exponential even though the distance dependence of the
individual components is not. The dependence upon the neutral-sphere
radius is dominated by the contribution from the exchange-correlation
part, which approximately scales as $s^2$, originating from the scaling of the
mean density in
the reference system.

\subsection{Calculating the one-electron correction with an LMTO tight-binding
model}
\label{sec:e1el}
Returning to the expression for the total energy, Eq.~(\ref{eq:etot}),
we have already given expressions for  the first two terms and need
only to calculate the one-electron correction to obtain the total
energy. This is  the most time
consuming step in a total-energy calculation
since it involves the diagonalization of a Hamiltonian.
The key number  in this context is the number of basis
functions ($N$) because the computer time used in conventional
diagonalization schemes scales as
O($N^3$)\footnote{In the Car-Parrinello method the scaling is
O($NM^2$) where M is the system size. However, the prefactor  is large
 such that this method generally is two orders of magnitude slower
than tight-binding methods.}.
A plane-wave basis set is in this respect not very efficient, since
many plane-waves are needed in order to get a good description of the
regions around the atomic
positions where the electron density varies rapidly. Instead, we will
use a partial
wave method in which the basis functions are augmented with the the
local solution
of the Schr\"{o}dinger equation around each atom, and therefore only a small
basis set is needed.

 However, the partial waves depend on the energy at which
the Schr\"{o}dinger equation is solved, and the Hamiltonian thereby
becomes energy
dependent. This problem is solved very elegantly in the linearized
band-structure methods where an energy independent basis set is
obtained by linearizing
the solutions of the Schr\"{o}dinger equation around a fixed energy
$\epsilon_{\nu}$. We will use the Linearized Muffin Tin Orbital (LMTO)
method\cite{tblmto}, and with  these  basis
functions, called LMTO's, the eigenvalues become correct
to first order in the difference with the energy $\epsilon_{\nu}$,
and a systematic expansion exists for the
 higher order corrections. For now we will only consider
the first order approximation, and we may then neglect the overlap of
the orbitals, since the overlap enters as a higher order correction.

Since we use the ASA for calculating the kinetic energy the first order
 LMTO Hamiltonian becomes especially simple.
It separates  into
potential parameters $\Delta^{\alpha}, C^{\alpha}$ determined
from the potentials in the atomic spheres, and structure constants $S^{\alpha}$
which only depend on the positions of the atomic spheres\cite{tblmto}
\begin{equation}
H_{iL,jL'}^{\alpha} =  C_{il}^{\alpha} \delta_{iL,jL'} \, + \,
(\Delta_{il}^{\alpha})^{\frac{1}{2}} S_{iL,jL'}^{\alpha}
( \Delta_{jl'}^{\alpha})^{\frac{1}{2}} \; ,
\label{eq:lmtoham}
\end{equation}
where we use the notation $L=lm$ for the angular-momentum quantum numbers,
and we will  use an $sp^3$ basis set.

The index $\alpha$ in Eq.~(\ref{eq:lmtoham}) denotes the representation
we use for the LMTO's. The conventional LMTO's are the $\alpha=0$
representation
in which the structure constants have simple two center forms
\begin{equation}
S^{0}_{iL',iL} = 0 \, , \; \;
S^{0}_{\{ss\sigma,sp\sigma,pp\sigma,pp\pi\}}  =  \{ -2 x^{-1},
2\sqrt{3} x^{-2},12 x^{-3} ,-6 x^{-3}\}\; ,
\label{eq:s0}
\end{equation}
where  $x$  is a relative distance measure given by
\begin{equation}
x =  d_{ij}/w_{ij} \, , \; \;  w_{ij} = s_{0} \frac{d[(s_i+s_j)/2]}{ d_0}.
\label{eq:x}
\end{equation}
In this equation $d[s]$ is a function that returns the nearest-neighbor
distance in the
diamond lattice with neutral-sphere radius $s$, such that we have
 the same relative distance $x_0 = d_{0}/s_{0}$, for all diamond lattices.
Note that the LMTO Hamiltonian will not depend
on the choice of $w$ as defined in Eq.~(\ref{eq:x}), since the $w$ dependence
of the
structure constants is canceled by a
similar term in the potential parameters.

It is possible to shift to a new representation
where the neighboring sites, through a  screening ``charge'' $\alpha$, are
used to localize the structure constants\cite{tblmto}.
The structure constants now
depend on the local structure through a matrix equation
(the LMTO Dysons equation)
\begin{equation}
S^{\alpha}_{iL,jL'} = S^{0}_{iL,jL'} + \sum_{kL''}
S^{0}_{iL,kL''} \alpha_{l''} S^{\alpha}_{kL'',jL'}.
\label{eq:lmtodys}
\end{equation}
We use the screening constants
$\alpha_{s,p} = \{0.3072,0.0316\}$ which are related to the $sp$-screening
parameters of Ref.~\cite{tblmto} through a scaling of $(1.07)^l$.

With this choice of screening constants we have calculated the structure
constants for all the test systems of Table~1 using the LMTO Dysons
equation, which involves inverting a 200x200 matrix.
In Fig.~\ref{fig:structconst} we show the Slater-Koster
components\cite{slakos}  of each structure matrix, as a function of the
relative distance
measure($x$) defined in Eq.~(\ref{eq:x}). It is surprising how well the
structure
constants for  such different surroundings as surfaces, phonons, and
different crystal structures, all fall on the same curves.
In Ref.~\cite{tblmto}
interpolation formulas for the structure constants of
close packed structures were found by using
a relative distance measure obtained by scaling the distances with the
Wigner-Seitz radius. The neutral-sphere provides a natural measure,
which makes it possible to extend this idea to more open structures
and surfaces.

With our choice of relative distance measure
we may now use the data base
to construct interpolation formulas and thereby obtain a simple and fast
scheme for calculating the  structure constants of a general system.
For the $ss\sigma$ and $pp\pi$ elements we only need to include
nearest-neighbor elements, while the $sp\sigma$ and $pp\sigma$ elements
are longer range, and we therefore have to include second-nearest neighbor
elements. For the interpolation we have used the functional form
\begin{eqnarray}
\label{eq:sfit}
\tilde{S}^{\alpha}(x) & = & f(x) - f(x_c) -(x-x_c) f'(x_c), \\
f(x) & = & A (1+ \lambda (x-1)) e^{-\lambda x}, \nonumber
\end{eqnarray}
where the first equation ensure that
the structure constants  go continuous differentiable to zero at
the cutoff $x_c $. The relative distance measure, $x$, is defined
in Eq.~(\ref{eq:x}) and $A,\lambda$
are parameters to be determined from the data base.
We fix the parameter $A$
 from the nearest-neighbor structure constant of the diamond lattice,
and determine $\lambda$
 by a least-squares fit to the nearest-neighbor data points of
Fig.~\ref{fig:structconst}. The resulting approximations are shown as
solid lines in Fig.~\ref{fig:structconst}, and the values of the parameters
are given in Table~\ref{tab:sfit}.

The  on-site elements of the screened structure matrixes are nonzero. We
have from  Eq.~(\ref{eq:lmtodys})
\begin{equation}
\sigma^{\alpha}_{iL,iL'} = \sum_{kL''} S^{0}_{iL,kL''} \alpha_{l''}
\tilde{S}^{\alpha}_{kL'',iL'},
\label{eq:sigma}
\end{equation}
where $\sigma^{\alpha}$ is the on-site element calculated using the approximate
structure matrix $\tilde{S}^{\alpha}$. For the diamond structure the on-site
element is diagonal and we have $\sigma^{\alpha}_{s,p}(diamond) =
\{1.71,1.46\}$,
while it will contain off-diagonal components in a general system.
However, the use of the approximate off-site structure
matrixes($\tilde{S}^{\alpha}$)
in Eq.~(\ref{eq:sigma}) gives a significant
error for these components, i.e. the off-diagonal components of
 $\sigma_{\alpha}$ are non symmetric and generally to large. We will therefore
use the following approximation for the on-site structure matrix
\begin{equation}
\tilde{S}^{\alpha}_{iL,iL'} = \frac{1}{4} [ \sigma^{\alpha}_{iL,iL'} +
\sigma^{\alpha}_{iL',iL}+2 \sigma^{\alpha}_{iL,iL'}(diamond) ],
\label{eq:sonfit}
\end{equation}
which is a simple average of the symmetrized value of $\sigma^{\alpha}$
between the system and the reference system.

We now return to the calculation of the potential parameters. These are
given by the solutions at the linearization energy $\epsilon_{\nu}$
of the radial Schr\"{o}dinger equation within each
atomic sphere\cite{skriver}.
Since we get the potential from the reference system,
we only have to calculate the potential parameters for the reference system
once and for all, and then use the neutral-sphere radius to
find the potential parameters for
a general system.
 In Fig.~\ref{fig:delta} and Fig.~\ref{fig:center}
we show the value of the potential parameters as a function
of the neutral-sphere radius of the reference system. For each system
we have chosen $\epsilon_{\nu}$ at the center of gravity
of the occupied bands. These data are accurately approximated by the
interpolation formulas
\begin{eqnarray}
\tilde{\Delta}_s^{\alpha}(s) & = &\Delta_s^{\alpha}(s_0)  e^{-1.130 (s
- s_0)} \; ,\\
\label{eq:delfits}
\tilde{\Delta}_p^{\alpha}(s) & = &\Delta_p^{\alpha}(s_0)  e^{-0.978 (s
- s_0)} \; , \nonumber \\
\tilde{C}_p - \tilde{C}_s & = & 4.67 +
(C_p^{\alpha}(s_0)-C_s^{\alpha}(s_0)-4.67) e ^{-0.76 (s
- s_0)} \; , \nonumber
\end{eqnarray}
where $\Delta_s^{\alpha}(s_0) ,\Delta_p^{\alpha}(s_0),C_s^{\alpha}(s_0)$ and
$C_p^{\alpha}(s_0)$ (unit eV) are  the
potential parameters in the equilibrium diamond lattice listed
in Table~\ref{tab:potdat}, and $s$ is the neutral-sphere (unit $a_0$).  In
Fig.~\ref{fig:delta}
and Fig.~\ref{fig:center} the interpolation formulas for the
potential parameters are  shown as solid lines.

We have now constructed the LMTO tight-binding Hamiltonian directly from
the data of the {\em ab initio} plane-wave calculation.
However, due to the small basis set and the incomplete description of
the interstitial region the calculated bandstructure for the
equilibrium diamond structure does not agree
completely with that of   the plane-wave calculation shown in
Fig.~\ref{fig:bandstruct} (SC).
 We find the occupied band to be 15 percent to wide and the band gap
at the $\Gamma$ point to be to small.
In order to improve the Hamiltonian we have made a least-squares fit of
 the potential parameters to the three lowest eigenstates at the $\Gamma$
point and the two lowest at the X point. By this procedure we
include the effect of the neglected orbitals in an indirect fashion.
 The resulting potential parameters are shown in
the second row of Table~\ref{tab:potdat}, and we see that
$\Delta_s^{\alpha}(s_0)$ is unchanged
while both $\Delta_p^{\alpha}(s_0)$ and $C_p^{\alpha}(s_0) - C_s^{\alpha}(s_0)$
have been rescaled with a factor $0.77$.
The corresponding bandstructure is shown in Fig.~\ref{fig:bandstruct} (EMTB),
and
we now have a good description of the occupied parts of the bands, and the
bandgap at the $\Gamma$ point.

We are now ready to calculate the one-electron correction ($\Delta
E_{1el}$) of Eq.~(\ref{eq:etot}). Instead of calculating  the band energy we
will
calculate the bond energy\cite{pettifor}, because we thereby remove any
first order dependencies on the onsite elements
\begin{equation}
E_{1el} = \sum_{k \in occ} \epsilon_k - \sum_i N_i {\cal E}_i.
\label{eq:bonde}
\end{equation}
In this equation $N_i$ is the site projected occupation and ${\cal
E}_i$ the site projected onsite element. The bond energy depends only
weekly on
the shift in the average on-site elements. For instance
for the three principal surfaces, the largest shift in the
average on-site element is at the (100) surface, where it is
0.84 eV  higher, than in the bulk.
Such a shift lowers the bond energy 0.19 eV compared to a calculation
where the average on-site element is fixed at the bulk value.
We have chosen to fix the average on-site element to be zero,
${\cal E}_i =0$, since by this approximation the second term in
Eq.~\ref{eq:bonde}
vanishes, and this greatly simplifies the calculation of forces.
Due to this approximation the total energy for the
 surfaces will be a little to high with the EMTB model.

In Table~\ref{tab:efite1el}
we show the one-electron correction obtained with this scheme
compared to the value of the plane-wave Harris functional calculation. We see
that
the correction generally is overestimated by 2 percent. We can improve
the values slightly by scaling  all potential parameters with
this factor, see Table~\ref{tab:potdat}, and we then obtain the one-electron
correction in the last column
of Table~\ref{tab:efite1el}.
We see that for all test systems the accuracy is acceptable when compared
with the value of the total energy of Table~\ref{tab:ediv}.
In the next section we will first sum up the ingredients of the EMTB, and
then compare it to empirical tight-binding schemes by applying both models
to various test systems.

\section{APPLICATIONS}
\label{sec:appl}
\subsection{The EMTB potential}
An EMTB calculation starts with  loading  precalculated values for
the functions $\Gamma(d,s),V(d,s)$ defined in Eq.~(\ref{eq:gamma})
and Eq.~(\ref{eq:vparpot}), and the atomic-sphere, $e_{as}^{ref}(s)$,
 and one-electron energy term, $e_{1el}^{ref}(s)$,
of the reference system. Next we calculate  the neutral-sphere radius
of each atom, $s_i$, from  Eq.~(\ref{eq:neuts}).  The total energy is given by
\begin{equation}
E = \sum_i e_c(s_i) + E_{as} - \sum_i e_{as}(s_i) +
   E_{1el} - \sum_i e_{1el}(s_i).
\end{equation}
The cohesive function $e_c(s)$ is defined in Eq.~(\ref{eq:unibind}), and
the atomic-sphere energy $E_{as}$ defined in Eq.~(\ref{eq:easv}). The
one-electron energy
is calculated from the Hamiltonian
\begin{equation}
\tilde{H}_{iL,jL'}^{\alpha} = \tilde{C}_{il}^{\alpha}(s_i)
\delta_{iL,jL'} + (\tilde{\Delta}_l^{\alpha}(s_i))^{\frac{1}{2}}
 \tilde{S}_{L,L'}^{\alpha}(x_{ij})
( \tilde{\Delta}_{l'}^{\alpha}(s_j))^{\frac{1}{2}} \; ,
\end{equation}
where the off-site structure constants are given by the interpolation
formula in Eq.~(\ref{eq:sfit}) with the parameters of
Table~\ref{tab:sfit}, and the relative distance measure ($x$) defined
in Eq.~(\ref{eq:x}). The on-site elements are given by
Eq.~(\ref{eq:sonfit},\ref{eq:sigma},\ref{eq:s0}), using the screening
$\alpha_{s,p} = \{0.3072,0.0316\}$.
The potential
parameters are calculated from the interpolation formulas of
Eq.~(\ref{eq:delfits})
with the constraint that the average onsite element at each site is
zero (${\cal E}_i =0$), and
using the values in the last row of Table~\ref{tab:potdat} for
$\Delta^{\alpha}(s_0),C^{\alpha}(s_0)$. With this model we have calculated the
total energy for the
test systems of Table~\ref{tab:etot} and the corresponding results are
shown in the third column.

\subsection{Empirical Tight-Binding (EmpTB)}
In a
conventional Empirical Tight-Binding  scheme the energy function
consists of an attractive band structure term describing the bonding in the
system and a
repulsive pair potential usually interpreted as  arising from
the overlap interaction. For the comparison we will use the
tight-binding model of Goodwin et. al\cite{pettb}, with a fixed cutoff
as in Ref.~\cite{hotb}, and in
the following we will denote this model EmpTB. The EmpTB is based on
 the nearest-neighbor tight-binding model of Harrison\cite{harrisontb},
in which the strength of the
hopping integrals  is obtained by fitting
the bandstructure, and the level splitting is taken to be identical to that
of the atom. Harrison assumed an universal decay of the pair potential
and hopping integrals, such that the only parameter remaining to be
determined is the strength of the pair potential, which was fixed to
give the correct equilibrium lattice constant. The resulting model
gives an excellent description of the elastic
properties of silicon  in the diamond structure, however the model
fails to predict the energies of different silicon phases. Goodwin
et. al made the model transferable to other structures by introducing
a scaling of the hopping integrals, and adjusting the level splitting.
Their model has four adjustable parameters
which were fixed to give the cohesive energy and bulk modulus of the
diamond and fcc structure. This scheme has proven very successful for
describing systems\cite{hotb} far from the structures where
the tight-binding parameters were fitted, and currently most empirical
tight-binding  schemes use a similar functional form.

The last column in Table~\ref{tab:etot} shows the total energies of the test
systems obtained with the EmpTB. We see that the elastic properties
are described very well by the EmpTB, while the total energies of
 the structures and surfaces  are not too accurate. However, note that
the energy of the crystal structures were
calculated at the equilibrium volume as obtained from  the
selfconsistent calculation (in Fig.~\ref{fig:phase} we show the full
phase diagram).

\subsection{Comparison between  the EMTB and EmpTB model.}
In Table~\ref{tab:etot} we show for the two models
 the hopping integrals and sp level splitting in the equilibrium
diamond lattice. For the EMTB we show the scaled parameters, and for
the EmpTB we have included the value of $\epsilon_p-\epsilon_s$ used in
Ref.~\cite{harrisontb}. In Fig.~\ref{fig:bandstruct} we show the bandstructure
of the two models compared to a selfconsistent calculation. For the EMTB
model the description of the occupied bands is excellent. In the
EmpTB model the description is reasonable, with largest error for
the lowest state at the $\Gamma$ point.
Also the bandgap at the $\Gamma$ point is too small in the EmpTB model, this is
of
importance when  the model is used together with the O(N) method  of
Ref.~\cite{orderN1,orderN2},
where a large bandgap is needed in order to make the scheme efficient.

The difference between the two tight-binding models become apparent when we
look at
the distance dependence of the matrix elements. In the EMTB the
distance dependence is divided into two parts, scaling of the  potential
parameters and the structure matrix. When we have a uniform
compression, without a structural change, only the potential
parameters change. These scale approximately as $(s_0/s)^3$,
which is
similar to the scaling of  Ref.~\cite{harrisontb}. When there is a
structural change the scaling should be stronger, and this scaling
enters through the structure constants. In the EmpTB these two basic
scalings are mixed into one function.

In the last column of Table~\ref{tab:efite1el} we
show the value of the one-electron correction as obtained when  the
EmpTB is used to describe the one-electron energy for the system and
reference system. Clearly
the EmpTB does not describe the one-electron correction, with largest
discrepancies for surfaces and structural energies. For these systems
we have found crystal field terms to be important, i.e. off diagonal on-site
elements and shifts of the level spacings. Such effects are not included in the
EmpTB, but enter in the EMTB
through Eq.~(\ref{eq:sigma}).

Besides the one-electron term there is a pair-potential term in both
models. However the EMTB pair potential is negative, while the EmpTB
pair potential is positive.
The EmpTB pair potential can therefore
not only describe the electrostatic and exchange-correlation
energy, but must include some terms from the kinetic energy. In the
work of Harrison\cite{harrisontb} the pair potential is viewed as an
overlap interaction, i.e. it is mainly due to the one-electron energy.

In the EMTB there is an additional term, the cohesive function.
The fact that this term is not equal to the sum of the reference
pair-potential and one-electron energies, illustrates the idea behind
the reference system: Because the reference system is chosen for a
given atom so that the environment of the atom is similar to the
environment in the real system it is possible to calculate the energy
difference with a rather crude tight-binding model. The largest error
is in the shift of the  average on-site term of the Hamiltonian, but since
the potentials are identical in the system and reference system this
error  for the system is canceled by a similar error in the one-electron
energy of the reference system, and  the correct binding energy is then
obtained through the cohesive function.

In the next section we will calculate the phase
diagram for silicon and in section~\ref{sec:100},~\ref{sec:111} we
will investigate some of the reconstructions of
 the (100) and (111) surfaces.

\subsection{Phase diagram}
In Fig.~\ref{fig:phase} we show the total energies versus volume for
the diamond, clathrate II\cite{clathrate}, $\beta$-tin, simple cubic, bcc and
fcc phases of
silicon calculated with the EMTB and EmpTB schemes.
The EMTB potential gives a good description of
 the cohesive energy and the equilibrium volume for all the phases
investigated, while the EmpTB predicts the
correct energy ordering of all the  phases, excluding the clathrate~II
structure which is lower than the diamond structure, but the
equilibrium volumes are shifted.   The small
cutoff of the interactions in the EmpTB model becomes visible for the
$\beta$-tin and fcc phase, i.e. second nearest neighbors enter within
the cutoff range. In the EMTB scheme we  also use a fixed cutoff, however in
this
case we use a relative distance measure which is scale-invariant, and
this ensures that for a given structure  we include the same number of
neighbors independent of the lattice constant.
 For the EMTB model we have found  crystal field effects to give an important
contribution
to the structural energies. For instance in the fcc phase
$\epsilon_p-\epsilon_s = 5.39 eV$ compared to
$\epsilon_p-\epsilon_s=5.87 eV$ in the diamond phase.

\subsection{The (100) surface}
\label{sec:100}

There has  been a large effort to understand the different reconstructions of
the
(100) surface, involving a wide range of experimental and
theoretical tools since  different reconstructions occur at
different length scales. However, the reconstructions are so complex
that there are still a lot of unsolved problems, especially for systems
involving too few atoms in order for elasticity theory to be correct,
and too many atoms to be feasible for plane-wave methods.

There is a general consensus that the main building block for all the
reconstructions is the buckled 1x2 dimer reconstruction. This structure was
first predicted by Chadi\cite{chadi} from a tight-binding calculation,
and lately also  selfconsistent plane-wave
calculations\cite{needs,dabrowski} have  found this
structure to be lower in energy than a symmetric dimer reconstruction.
So for a potential to give
a detailed description of the various  reconstructions on the (100) surface,
it should at least predict the correct 1x2 dimer reconstruction.

In Table~\ref{tab:surfcalc} we show the result of a relaxation of the
(100) 1x1 and 2x1 reconstructions using the EMTB and EmpTB compared to the
results of  selfconsistent plane-wave calculations.
For both models the description of the structure of the 1x2 reconstruction is
good, but the EMTB gives a too high relaxation energy, while the
the relaxation energy of the EmpTB is far too small. For the EMTB the
driving force for the reconstruction is the atomic-sphere correction,
and not the one-electron correction. The atomic-sphere correction
always drives the system to more close packed structures, and in
this case that can be done without an increase in the one-electron energy.
Both models predict an outward relaxation of the  1x1 cell, while
the selfconsistent calculation predicts an inward relaxation.

\subsection{The (111) surface}
\label{sec:111}

Also the (111) surface has many interesting reconstructions,  the
most famous being the 7x7 Tagayanagi reconstruction\cite{tagayanagi}. This
reconstruction is built from adatom geometries where the adatoms sit on
top sites, and studies using selfconsistent plane-wave calculations
have confirmed the stability of adatoms in top site
positions\cite{northrup,adatom}. For a potential to predict the
correct reconstructions of the (111) surface it therefore has to describe these
adatom geometries properly.

In Table~\ref{tab:surfcalc} we show the result of a relaxation of the
(111) 1x1 surface, and of the top and hollow site adatom geometries in
a $\sqrt{3}X\sqrt{3}$ cell.
The selfconsistent numbers are from two
different references; The top site geometry were calculated in
 Ref.~\cite{adatom} with a ten layer slab and a 12
Ry cutoff. In the other reference\cite{northrup} the energy of the adatom in
both the top and
hollow positions where calculated but with an eight layer slab using a
6 Ry cutoff.  We see that both tight-binding models predict
the top site to be more stable than the hollow site geometry, however
the formation energies obtained with the EmpTB are far too small, while
the EMTB is in excellent agreement with the data of Ref.~\cite{adatom}.
The formation energy obtained with the  empirical tight-binding scheme
is too low to stabilize the adatom, and therefore
additional parameters have to be introduced in order to describe the
 7x7 reconstruction\cite{cha111}.
In Fig.~\ref{fig:T4} we show the geometry of the adatom in the top
position, and in Table~\ref{tab:surfcalc} we compare the relaxed coordinates.

\section{SUMMARY AND CONCLUSIONS}

In the present paper we have presented a new method for total energy
calculations for Si systems. We have discussed in detail the hierarchy of
approximations behind the present formulation. Starting from a fully
selfconsistent solution of the Kohn-Sham equations the first level of
approximation is to use the Harris functional with transferable,
optimized densities centered at each atomic position deduced from
independent selfconsistent calculations for different bulk and surface
structures. At this level the errors introduced compared to the fully
selfconsistent calculations are very small.

The next level of approximation involves the introduction of a
reference system. We have shown that choosing a reference system with
the same neutral-sphere radius (or average electron density) as in the
real system gives one-electron potentials and potential parameters for
the LMTO Hamiltonian that can be transferred from the reference system
to the real system with very little error. The energy difference
between the reference system and the real system can be calculated
from a difference between a density-dependent pair-potential sum in
the real and the
reference system and a one-electron energy difference. The former
describes to a very good approximation the difference in the
electrostatic, exchange correlation and part of the kinetic energy.
The one-electron energy difference taking care of the rest
of the energy difference is evaluated using an LMTO tight-binding
Hamiltonian.

The results of the new method are very encouraging. The computational
effort is similar to empirical tight-binding methods but
the results seem to be better. The time consuming part of the
calculation is the diagonalization of the LMTO Hamiltonian and for
this part the newly proposed "order N" methods can be
used\cite{orderN1,orderN2}, since the bandgap at the $\Gamma$ point is
well described.

\acknowledgements

We are grateful to H. Skriver whose LMTO programs we have used for calculating
the LMTO potential parameters.
We would also like to thank K. Kunc, O.H. Nielsen, R.J. Needs, and R.M. Martin
whose solid state programs we have used, and to E.L. Shirley who developed
the pseudo-potential routines. This work was in part supported by the Danish
Research Councils through the Center for Surface Reactivity. The Center for
Atomic-scale Materials Physics is sponsored by the Danish National Research
Foundation. Nithaya Chetty acknowledges
support from the Division of Materials Sciences, U.S.
Department of Energy under contract No. DE--AC02--76CH00016.

\appendix
\section{Parameterization of the optimized density}

\begin{equation}
\Delta n_{op}(k) \,=\, \Delta n_{atom}(k) +
a_{1}k^{2}(k-a_{2})(k-a_{3}) \times
exp(a_{4}k-a_{5}k^{2}),
\end{equation}

where $a_{1} = 0.078, a_{2} = 2.584, a_{3} = 1.465, a_{4} = 1.969,
a_{5} = 0.962$ and $n_{atom}$ is the atomic density.

\newpage

\begin{figure}
\caption{The free atom density (solid) and optimized density (dashed)
for Si. \label{fig:opdens}}
\end{figure}

\begin{figure}
\caption{The neutral-sphere radius(solid) for silicon in the diamond lattice
as a function of the nearest neighbor distance. The dashed lines show the
Wigner-Seitz radius of the diamond lattice, and of an inscribed bcc
lattice. \label{fig:dsss} }
\end{figure}

\begin{figure}
\caption{The selfconsistent potential of silicon in the diamond lattice
projected onto the $(1\bar{1}0)$ plane. The solid circle show the radius of
the neutral-sphere radius, while the dashed circle show the Wigner-Seitz
radius. \label{fig:sipot}}
\end{figure}

\begin{figure}
\caption{The density-dependent pair potential, and its three components, used
for
calculating the atomic-sphere correction as function of distance, with
the  neutral-sphere radius fixed at $s_0$. The inlet
shows the logarithm of the pair potential. \label{fig:vd}  }
\end{figure}

\begin{figure}
\caption{The pair potential as a function of the neutral-sphere radius
($s$), with the distance fixed at $d_0$. \label{fig:vs}  }
\end{figure}

\begin{figure}
\caption{The crosses show the Slater-Koster components of the structure
constants for
the structures in Table~\protect\ref{tab:etot} as a function of the
relative distance measure, $x$, defined in Eq.~(\protect\ref{eq:x}).  The
structure constants were calculated using
Eq.~(\protect\ref{eq:lmtodys}). The tick marks on the horizontal axis
indicate the relative distance in the diamond, simple cubic, fcc and
bcc structure. The solid lines show the value of the structure
constant as obtained from  the interpolation formula of
Eq.~(\protect\ref{eq:sfit}).\label{fig:structconst} }
\end{figure}

\begin{figure}
\caption{The dots  show the calculated potential parameters
$\Delta^{\alpha}$ for the diamond reference system, the solid line
is the approximation obtained with the interpolation formulas
 of Eq.~(\protect\ref{eq:delfits}). \label{fig:delta}  }
\end{figure}

\begin{figure}
\caption{The dots show the potential parameters $C^{\alpha}$, and the
solid line is the approximation obtained with the interpolation formulas
 of Eq.~(\protect\ref{eq:delfits}). \label{fig:center} }
\end{figure}

\begin{figure}
\caption{The figure show the bandstructure of silicon calculated with
three different methods. Starting from the top the calculations are:
Self-consistent plane-wave calculation (SC),  the EMTB model and the
Empirical Tight-Binding model(EmpTB) of Ref.~\protect\cite{pettb}.
\label{fig:bandstruct} }
\end{figure}

\begin{figure}
\caption{The triangles in both figures show the energies of the diamond,
clathrate~II,
$\beta$-tin, simple cubic, bcc and fcc structure, in that order,
at their respective
equilibrium volumes, calculated selfconsistently.
The energy of the clathrate~II structure is from
Ref.~\protect\cite{clathrate}. The solid lines in the upper figure show the
energies of the structures calculated using the EMTB model, while the
energies in the lower figure where calculated using the empirical
tight-binding model of Ref.~\protect\cite{pettb}. The cusps on the curves in
the lower figures are caused by second-nearest neighbors entering
within the cutoff distance. \label{fig:phase} }
\end{figure}

\begin{figure}
\caption{The geometry of the top site adatom on the (111) surface.
\label{fig:T4} }
\end{figure}

\newpage

\mediumtext
\begin{table}
\caption{The selfconsistent (SC), the Harris functional (H) and the
Effective-Medium Tight-Binding (EMTB) results for
the lattice constant $a_0$, bulk modulus $B$ and cohesive energy $E_c$
for Si in the diamond structure. The energy relative to the diamond
phase of the $\beta$-tin, simple cubic, bcc
and fcc structure at the equilibrium lattice constant as obtained from  the
selfconsistent calculation.   The phonon frequencies of the transverse
acoustics phonon at the $X$
point, the transverse optical phonon at the $X$
point,  the longitudinal acoustic
and optical phonon at the $X$ point $LAO(X)$, and the longitudinal and
transverse optical
phonon at the $\Gamma$ point $LTO(\Gamma)$. The three cubic elastic constants.
The energies of the
ideal principal surface orientations. In the last column (EmpTB) we show
the values obtained with the empirical tight-binding model of
Ref.~\protect\cite{pettb}.\label{tab:etot} }

\begin{tabular}{lcccc|c}
\multicolumn{1}{c}{Quantity}   &
\multicolumn{1}{c}{units}      &
\multicolumn{1}{c}{SC}       &
\multicolumn{1}{c}{H}        &
\multicolumn{1}{c}{EMTB}     &
\multicolumn{1}{c}{EmpTB}      \\
\tableline
\multicolumn{4}{l}{a) DIAMOND BULK }  \\
$a_0$            & \AA  & 5.395       & 5.380 & 5.380   & 5.42       \\
$B$              & MBar & 0.99        & 0.93  & 0.93    & 1.04       \\
$E_c$            & eV   & 5.85        & 5.83  & 5.83    & 4.70        \\
\multicolumn{4}{l}{b) STRUCTURES} \\
$\beta$-tin (4.76 \AA) & eV/atom & 0.26 & 0.25 & 0.27  &   0.60\\
sc(2.51 \AA) & eV/atom & 0.41 & 0.41 & 0.43 & 0.73 \\
bcc(3.03 \AA)& eV/atom & 0.55 & 0.52 & 0.73 & 1.31 \\
fcc(3.79 \AA) & eV/atom & 0.56 & 0.54 & 0.89 & 1.32  \\
\multicolumn{4}{l}{c) PHONONS} \\
 $\nu_{TA}(X)$ & THz &  4.3 & 4.2 & 4.0  &  5.0\\
 $\nu_{TO}(X)$ & THz &  14.1 &  13.3 & 19.0 &  16.6 \\
 $\nu_{LOA}(X)$ & THz &  12.3 &  12.2 & 12.7 &  14.4 \\
 $\nu_{LTO}(\Gamma)$ &THz &  15.7 &  15.6 & 19.6 & 18.5\\
\multicolumn{4}{l}{d) ELASTIC CONSTANTS} \\
 $C_{11}$ &Mbar &  1.70 & 1.65 & 1.43 &  1.46	\\
 $C_{12}$ &Mbar  &  0.72 &  0.62& 0.90 & 0.78 \\
 $C_{44}^{0}$ &Mbar  &  1.10&  1.10 & 1.69 & 1.23  \\
 \multicolumn{4}{l}{e) SURFACES} \\
 (100) &eV /atom     &2.19   &  2.10  &  2.39 & 1.65  \\
  (110) & eV/atom     &  1.27  &  1.28  &  1.55 & 1.43 \\
 (111) & eV /atom    &  1.45  &  1.43  &   1.56 & 1.43 \\
\end{tabular}
\end{table}

\begin{table}
\caption{The effective-medium components of the Harris functional results of
Table~\protect\ref{tab:etot}. The second column show the lattice
constant of the structure and the displacement used for the frozen
phonon and  elastic deformation calculations.\label{tab:ediv} }
\begin{tabular}{lcl|ccccc}
& & & \multicolumn{1}{c|}{$E_c$}        &
\multicolumn{3}{c|}{$\Delta E_{as}$}      &
\multicolumn{1}{c}{$\Delta E_{1el}$}      \\
\multicolumn{1}{l}{Quantity} &  & \multicolumn{1}{c|}{H}       &
 \multicolumn{1}{c|}{} & $\Delta G_{el}$ & $\Delta G_{xc}$ &
\multicolumn{1}{c|}{$\Delta V$} \\
\tableline
$\beta$-tin & (4.76 \AA) &  0.25 & 0.00 & -1.39 & -0.80 & 0.67 & 1.78 \\
sc & (2.51 \AA)  &  0.41 & 0.01 & -0.99 & -0.64 & 0.37 & 1.67 \\
bcc & (3.03 \AA) &  0.52 & 0.01 & -1.94 & -1.13 & 0.96 & 2.62 \\
fcc & (3.79 \AA)  & 0.54 & 0.00 & -1.75 & -1.16 & 0.61 & 2.84 \\
 TA(X) & (0.04) &  0.025 & 0.002 &  -0.053 & -0.022 & 0.035 & 0.062  \\
 TO(X) & (0.02)  & 0.059 &  0.001 & -0.062 & 0.023 & 0.014 & -0.044 \\
 LOA(X) & (0.02) & 0.099 &  0.029 & -0.050 & 0.007 & 0.096 & 0.017 \\
 LTO$(\Gamma)$ & (-0.01) & 0.145  &  0.002 & -0.133 & 0.031 & 0.277 & -0.032 \\
 LTO$(\Gamma)$ & (0.01) & 0.099  &  0.001 & -0.068 & 0.025 & 0.150 & -0.009 \\
 $C_{11}$ & (0.06) & 0.0361 & 0.0229 & -0.0060 & -0.0034 & 0.0051 & 0.0174 \\
 $2(C_{11} -C_{12})$ & (0.06) & 0.0450 & 0.0005 & -0.0257 & -0.0117 & 0.0170 &
0.0650 \\
$C_{44}^{0}$ & (0.06)& 0.0241 & 0.0003 & -0.0235 & 0.0050 & 0.0463 & -0.0039 \\
 (100)     &  & 2.10   & 0.40 & 0.36 & 0.68 & 0.32 & 0.34   \\
  (110)    & &  1.28   & 0.10 & 0.20 & 0.52 & 0.23 & 0.23   \\
 (111)     & &   1.43  & 0.10 & 0.24 & 0.58 & 0.24 & 0.27   \\
\end{tabular}
\end{table}

\begin{table}
\caption{The table shows the value of the $\Delta G_{xc}$  term for the
structures of
Table~\protect\ref{tab:ediv}. The column denoted H is the value  of
the term calculated using the Harris functional, the  column denoted
EMTB is the result obtained using the approximations of the EMTB
model, and in  the third column a parameter has been allowed in order to
scale all the energies. \label{tab:efitxc}}
\begin{tabular}{lc|ccc}
& &  \multicolumn{3}{c}{$\Delta G_{xc}$}   \\
\multicolumn{1}{l}{Quantity} &  & H & EMTB & x$1.28$ \\
\tableline
$\beta$-tin & (4.76 \AA) &  -0.80  & -0.62  & -0.80  \\
sc & (2.51 \AA)          &  -0.64  & -0.48  & -0.62  \\
bcc & (3.03 \AA)         &  -1.13  & -0.90  & -1.16  \\
fcc & (3.79 \AA)         &  -1.16  & -0.94  & -1.21   \\
 TA(X) & (0.04)          &  -0.022 & -0.015 & -0.020  \\
 TO(X) & (0.02)          &  0.023  & 0.016  & 0.020   \\
 LOA(X) & (0.02)         &  0.007  & 0.001  & 0.002   \\
 LTO$(\Gamma)$ & (-0.01)  &  0.031  & 0.023  & 0.029  \\
 LTO$(\Gamma)$ & (0.01)  &  0.025  & 0.018  & 0.023   \\
 $C_{11}$ & (0.06)       &  -0.0034&-0.0025 & -0.00339 \\
 $2(C_{11} -C_{12})$ & (0.06)& -0.0117 & -0.0082 & -0.0105 \\
$C_{44}^{0}$ & (0.06)    &  0.0050 & 0.0030 & 0.0038 \\
 (100)     &             &  0.68   & 0.31   & 0.40   \\
  (110)    &             &  0.52   & 0.25   & 0.32   \\
 (111)     &             &  0.58   & 0.26   & 0.33   \\
\end{tabular}
\end{table}

\begin{table}
\caption{The value of the parameters in Eq.~(\protect\ref{eq:sfit}).
The parameter $x_c$ determines the range of the Hamiltonian, which
for the $ss\sigma$ and $pp\pi$ element is nearest neighbor in the diamond
lattice,
while it is second nearest neighbor for the $sp\sigma$ and $pp\sigma$ element.
The $S^{\alpha}(d_0/s_0)$ is the structure constants of the equilibrium diamond
lattice as obtained from the LMTO Dysons equation, while $\lambda$ is
obtained from a least squares fit to the data points of
Fig.~\protect\ref{fig:structconst}. \label{tab:sfit}}
\begin{tabular}{l|cccc}
quantity & ss$\sigma$ & sp$\sigma$ & pp$\sigma$ & pp$\pi$ \\
\hline
$x_c$ & 2.60 & 2.95 & 2.95 & 2.60 \\
$S^{\alpha}(d_0/s_0)$ & -0.938 & 1.690 & 3.279 & -1.025   \\
$\lambda$ & 2.40  & 2.85 & 2.76 & 4.10 \\
\end{tabular}
\end{table}

\begin{table}
\caption{The LMTO potential parameters for the equilibrium diamond lattice.
The values in the first row are obtained by solving the radial Schr\"{o}dinger
equation within the atomic-sphere, the values in the second row are
obtained by a least squares fit to the bandstructure of silicon, and
the values in the last row are those of the second row but
 rescaled with 0.98.\label{tab:potdat}}
\begin{tabular}{l|ccc}
quantity & $C_p^{\alpha}(s_0)-C_s^{\alpha}(s_0)$ & $\Delta_s^{\alpha}(s_0)$ &
$\Delta_p^{\alpha}(s_0)$ \\
\tableline
calc. & 10.73 & 1.954 & 0.959 \\
fit  &  7.91  & 1.954 & 0.738 \\
x0.98 & 7.75   & 1.915 & 0.732 \\
\end{tabular}
\end{table}

\begin{table}
\caption{The table shows the value of
$\Delta E_{1el}$ terms for the structures of
Table~\protect\ref{tab:ediv}. The column denoted H is the value  of
the term calculated using the Harris functional, the  column denoted
EMTB is the result obtained using the approximations of the EMTB
model, and in  the third column a parameter has been allowed in order to
scale all the energies. The last column (EmpTB) shows
the values obtained with the empirical tight-binding model of
Ref.~\protect\cite{pettb} \label{tab:efite1el}}
\begin{tabular}{lc|ccc|c}
& & \multicolumn{4}{c}{$\Delta E_{1el}$}      \\
\multicolumn{1}{l}{Quantity} &  & H & EMTB  & x$0.98$ & EmpTB \\
\tableline
$\beta$-tin & (4.76 \AA) & 1.78    & 1.81   & 1.77  & 0.38 \\
sc & (2.51 \AA)          & 1.67    & 1.69   & 1.66  & 0.81 \\
bcc & (3.03 \AA)         & 2.62    & 2.90   & 2.84  & 0.29 \\
fcc & (3.79 \AA)         & 2.84    & 3.29   & 3.23  & 0.44 \\
 TA(X) & (0.04)          & 0.062   & 0.058  & 0.057 & 0.057 \\
 TO(X) & (0.02)          & -0.044  & 0.022  & 0.022 & 0.049 \\
 LOA(X) & (0.02)         & 0.017   & 0.031  & 0.031 & 0.062 \\
 LTO$(\Gamma)$ & (-0.01) & -0.032  & 0.053  & 0.052 & 0.1081 \\
 LTO$(\Gamma)$ & (0.01)  & -0.009  & 0.055  & 0.054 & 0.1233 \\
 $C_{11}$ & (0.06)       & 0.0174  & 0.0162 & 0.0159& 0.0137 \\
$2(C_{11} -C_{12})$&(0.06)& 0.0650 & 0.0530 & 0.0520& 0.0422 \\
$C_{44}^{0}$ & (0.06)    & -0.0039 & 0.0126 & 0.0123& .0168 \\
 (100)     &             & 0.34    & 0.91   & 0.89  & 1.15\\
  (110)    &             & 0.23    & 0.68   & 0.67  & 1.23 \\
 (111)     &             & 0.27    & 0.65   & 0.64  & 1.23 \\
\end{tabular}
\end{table}

\begin{table}
\caption{The tight-binding parameters of the equilibrium silicon
structure for the EMTB and the Empirical Tight-binding (EmpTB) scheme of
Ref.~\protect\cite{pettb}. For the EMTB we show the rescaled
parameters. In  Ref.~\protect\cite{pettb} the four hopping integrals for the
EmpTB were taken from
Ref.~\protect\cite{harrisontb} and the level splitting  fitted
to the fcc -- diamond energy difference. The number in parenthesis is
the value of Ref.~\protect\cite{harrisontb}.\label{tab:comp} }
\begin{tabular}{l|ccccc}
model &$ H_{ss\sigma} $& $H_{sp\sigma}$ & $H_{pp\sigma}$ &
$H_{pp\pi}$ & $\epsilon_p-\epsilon_s$\\
\hline
EMTB & -1.79 & 1.99  & 2.37 & -0.74 &  5.87 \\
EmpTB & -1.82 & 1.96 & 3.06 & -0.87 & 8.295(6.83) \\
\end{tabular}
\end{table}

\begin{table}
\caption{Energies and structures of the (100) and (111) surface
obtained from selfconsistent, EMTB, and EmpTB calculations.
The $\gamma$ denote the surface energy per 1x1 cell, and $\Delta \gamma$
 the surface energy relative to the ideal 1x1 cell. For the relaxed
1x1 geometries we show the relative relaxation of the first layer
atoms, $\Delta_{12}$. For the (100) 1x2 structure we show the dimer
bond length, $r_d$, and buckling angle, $\theta$. For the (111)
$\protect\sqrt{3}X\protect\sqrt{3}\; T_4$  structure we show the relaxation
toward the
adatom axis (dotted line in Fig.~\protect\ref{fig:T4}), $\delta r$,
and the relaxation in the vertical direction, $\delta z$.
The atom numbers  refer to Fig.~\protect\ref{fig:T4}.\label{tab:surfcalc} }
\begin{tabular}{llcccc}
\multicolumn{1}{l}{geometry}   &
\multicolumn{1}{c}{quantity}      &
\multicolumn{1}{c}{unit}      &
\multicolumn{1}{c}{SC}   &
\multicolumn{1}{c}{EMTB}   &
\multicolumn{1}{c}{EmpTB}  \\
\tableline
\multicolumn{6}{l}{THE (100) SURFACE}\\
Ideal 1x1 & $\gamma$ & eV & 2.19& 2.39 & 1.65 \\
Rel. 1x1  &  $\Delta \gamma$  & eV & $-0.03^a$ & -0.02 & -0.03 \\
          & $\Delta_{12} $ & \% & $-5.1^a$ & 4.8 & 4.5 \\
Rel. 1x2  &  $\Delta \gamma$  & eV& $-0.85^b$ & -1.04 & -0.39 \\
          & $\theta$ & Degree & $(16^{0})^{b}$ & $19^0$ & $15^0$\\
          & $r_d $  & $a_0$ & $4.28^b$ & 4.47 & 4.62 \\
\hline
\multicolumn{6}{l}{THE (111) SURFACE}\\
Ideal 1x1 & $\gamma$  & eV& 1.45 & 1.56 & 1.43  \\
Rel. 1x1  &  $\Delta \gamma$  & eV& $-0.06^c$  $(-0.17^d)$ & -0.01 & -0.02 \\
	& $\Delta_{12} $ & \% & $-27^c$ & -7 & 3 \\
$\sqrt{3}X\sqrt{3}\; T_4$  &  $\Delta \gamma$ & eV & $-0.27^c$ $(-0.28^d)$ &
-0.27 & -0.04
\\
&  $r_d $         & $a_0$  & $5.01^c$ $(4.70^d)$  & 4.89  & 4.95  \\
&  $\delta r (2)$ & $a_0$  & $-0.35^c$ $( -0.28^d)$ & -0.35 & -0.23 \\
&  $\delta z (2)$ & $a_0$  & $0.03^c$ $( -0.15^d)$ & 0.10  & 0.12 \\
&  $\delta z (3a)$ & $a_0$ & $-0.71^c$ $( -0.74^d)$ & -0.65 & -0.57 \\
&  $\delta z (3b)$ & $a_0$ & $0.38^c$ $( 0.18^d)$  & 0.18  & 0.21 \\
&  $\delta z (4a)$ & $a_0$ & $-0.54^c$ $( -0.48^d)$ & -0.58 & -0.62 \\
&  $\delta z (4b)$ & $a_0$ & $0.23^c$ $(  0.11^d)$ & 0.06  & 0.12 \\
&  $\delta r (5)$ & $a_0$  & $0.10^c$  & 0.09  & 0.13 \\
&  $\delta z (5)$ & $a_0$  & $0.01^c$  & 0.03  & -0.04 \\
$\sqrt{3}X\sqrt{3}\;  H_3$ &  $\Delta \gamma$ & eV & $-0.17^d$ & -0.13 & -0.02
\\
\end{tabular}
$^a$Reference~\protect\cite{yincohen}
$^b$Reference~\protect\cite{dabrowski}
$^c$Reference~\protect\cite{adatom}
$^d$Reference~\protect\cite{northrup}

\end{table}

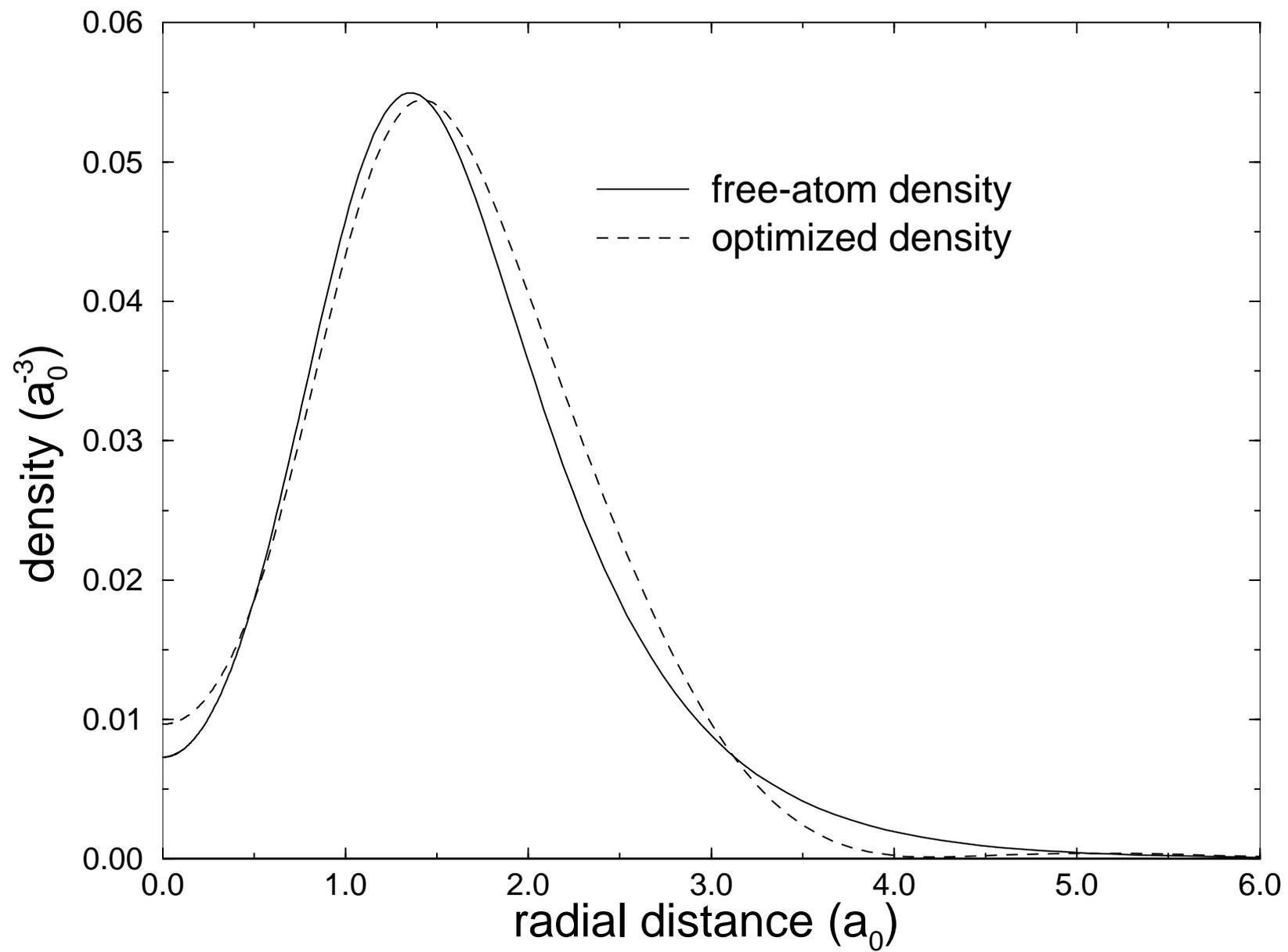
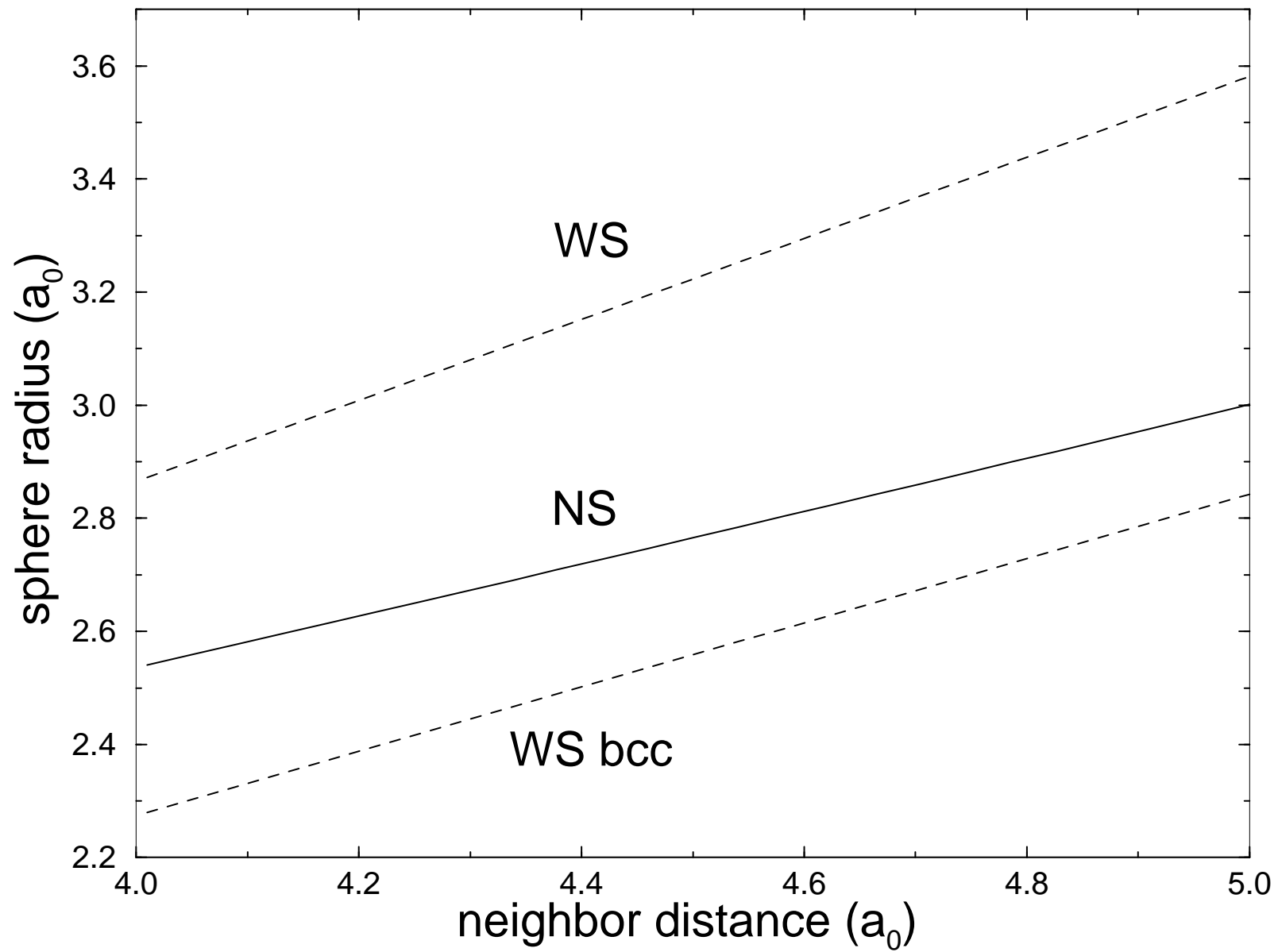
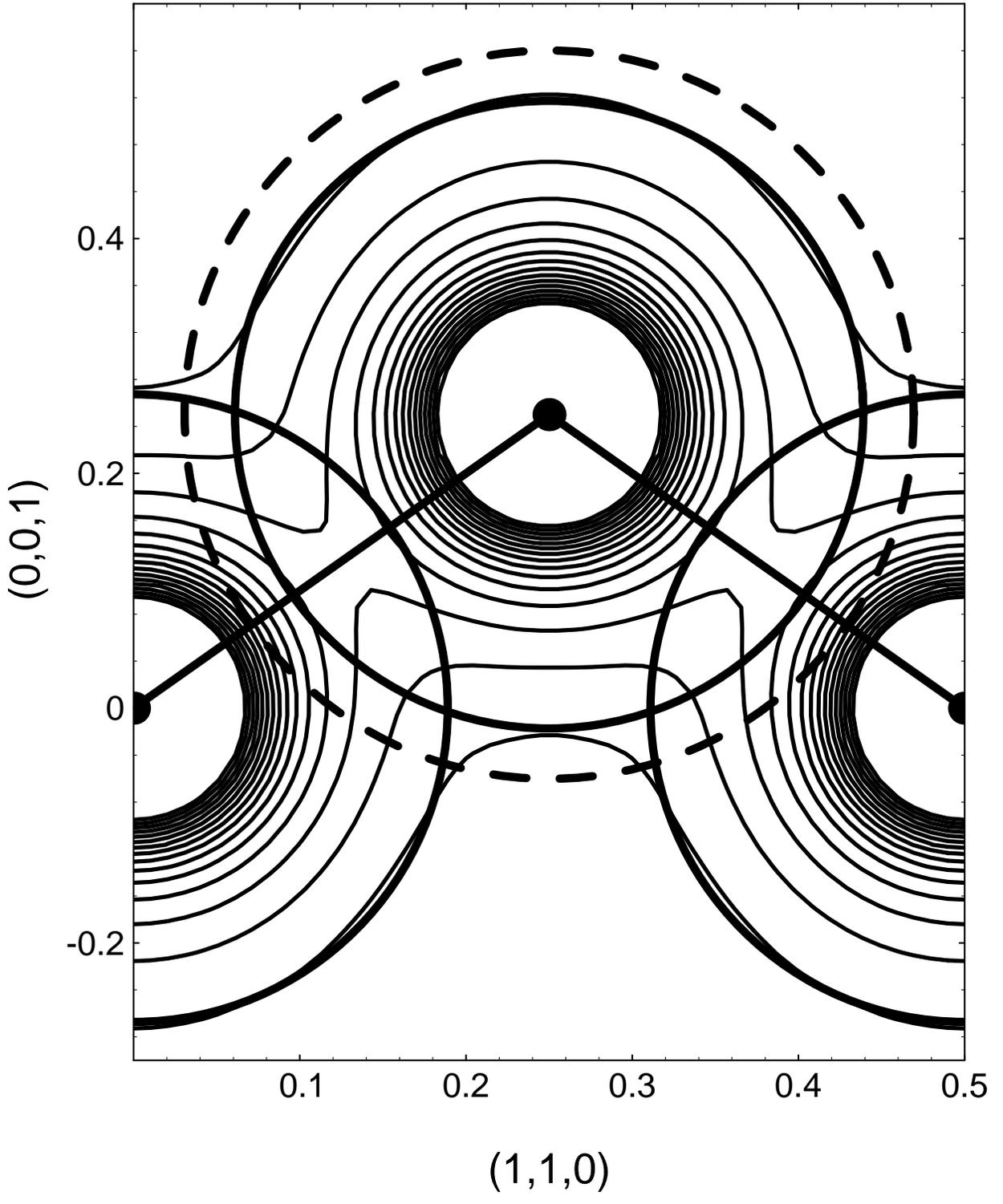
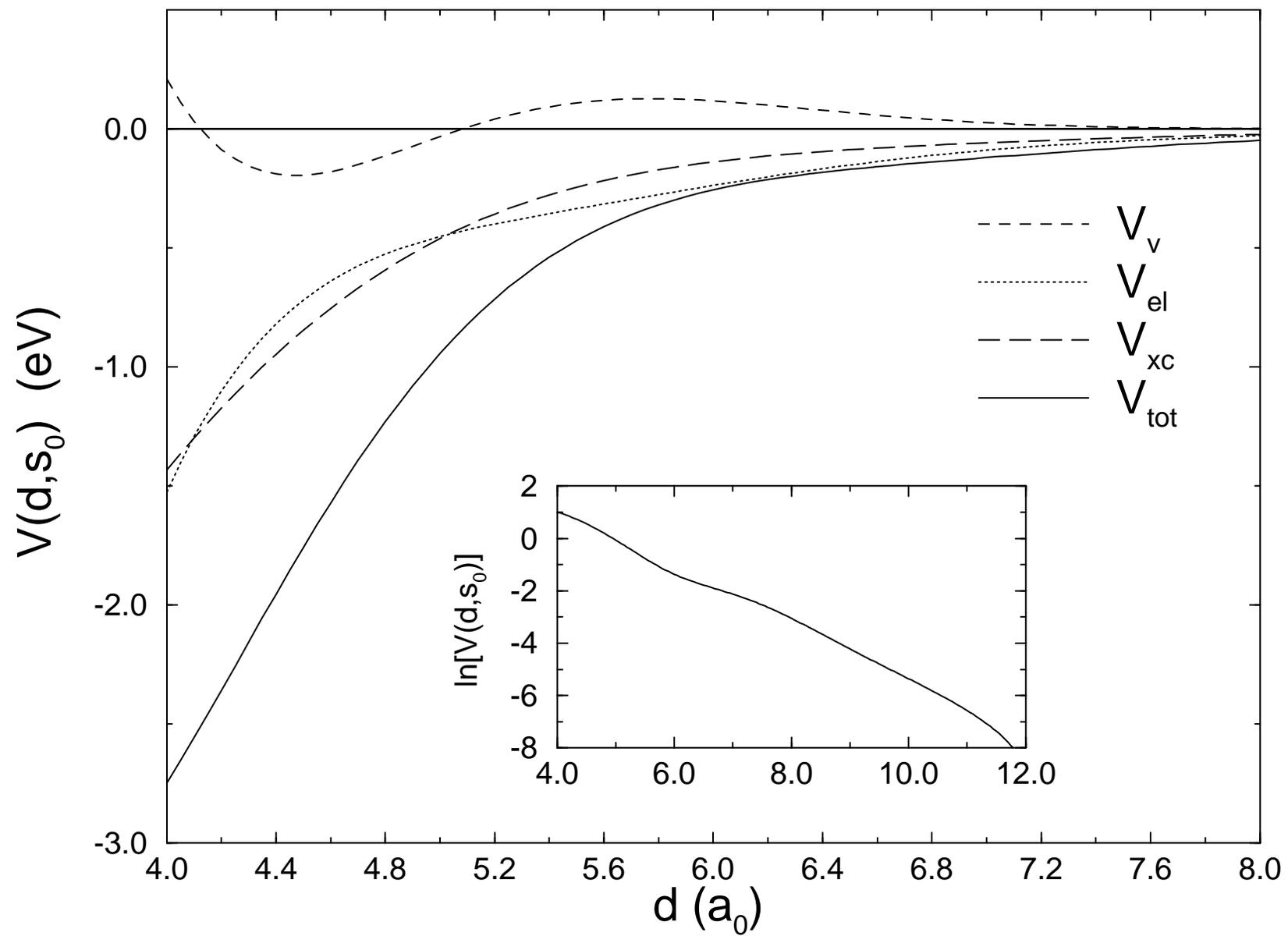
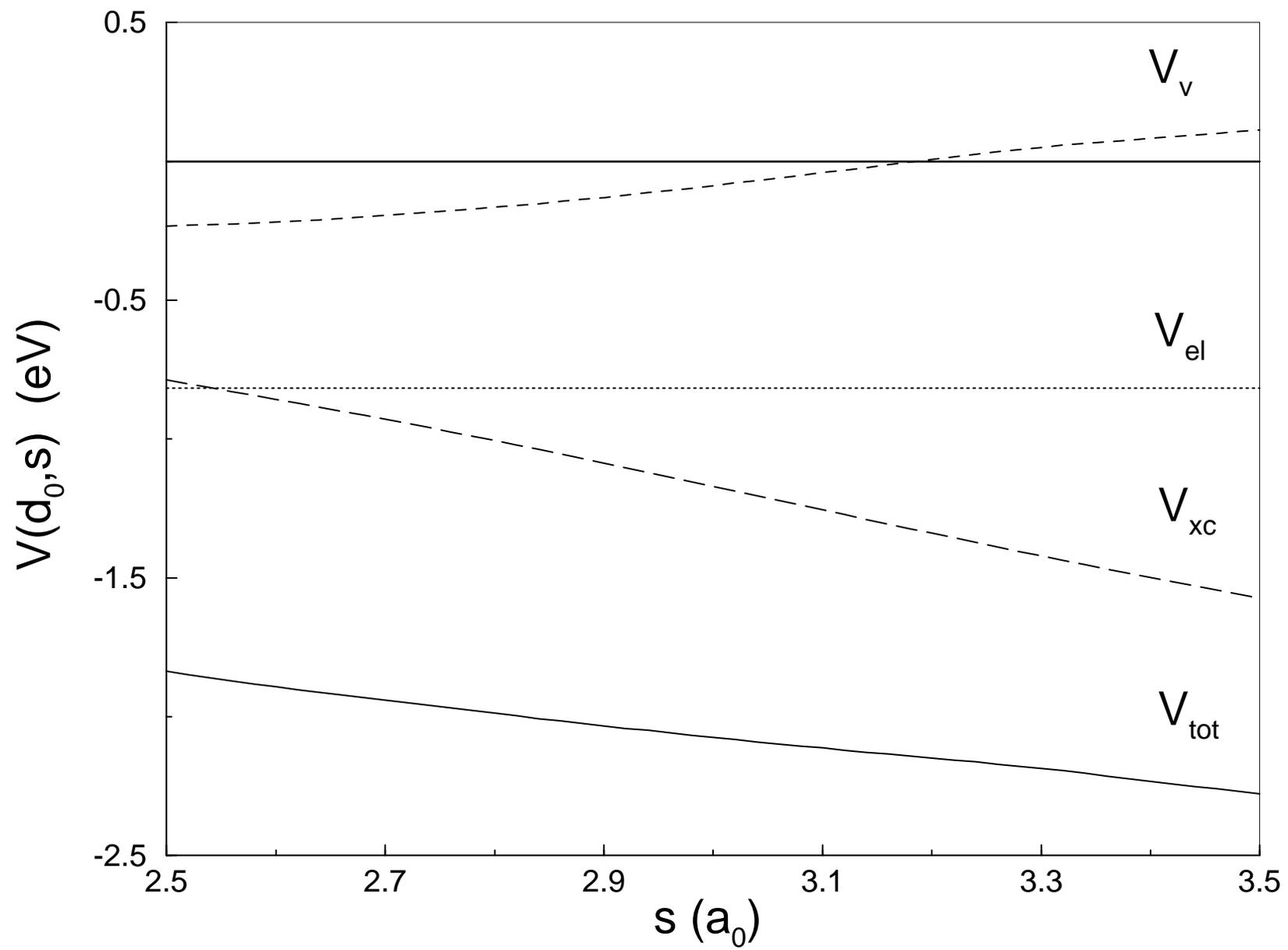
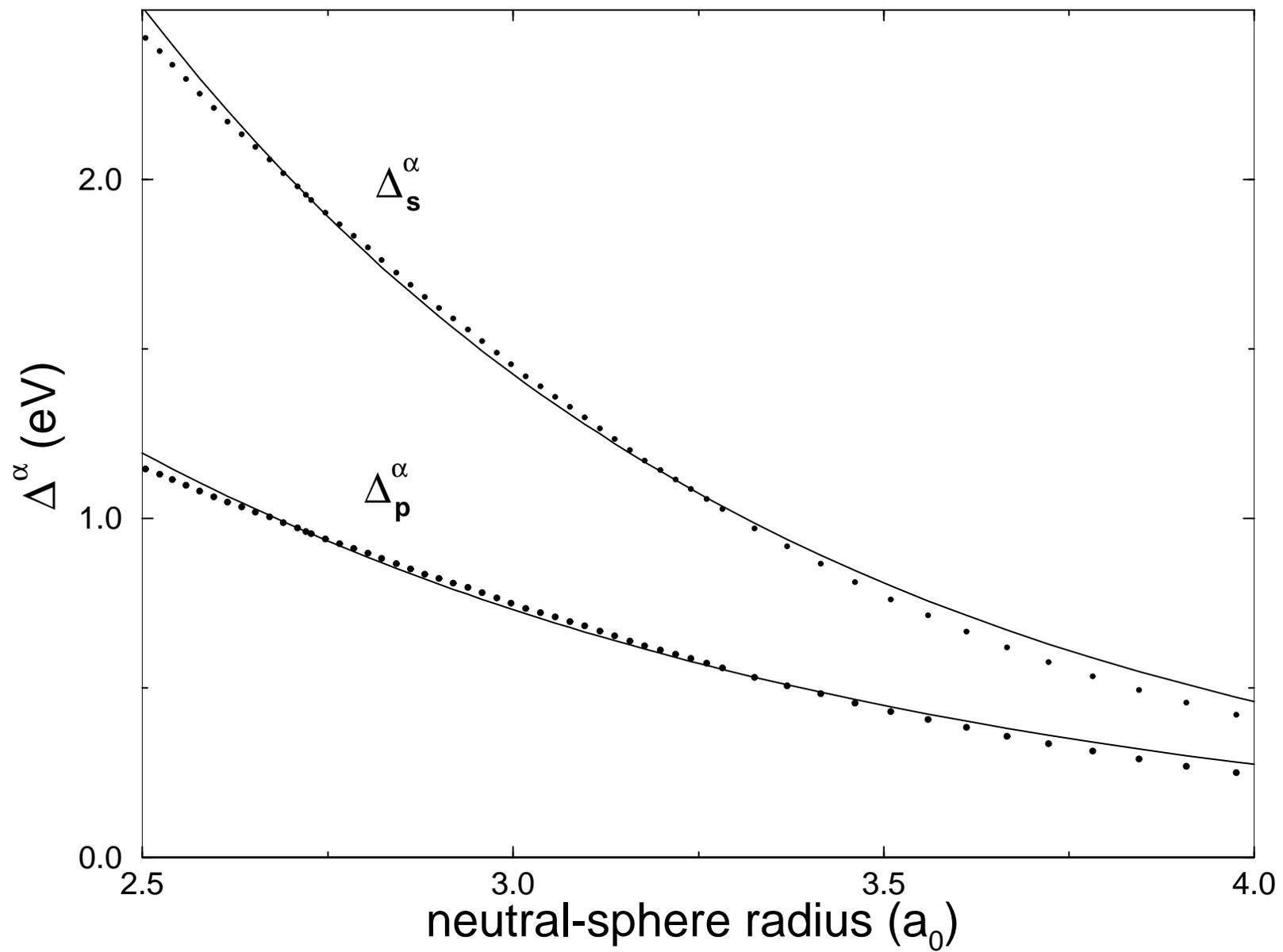
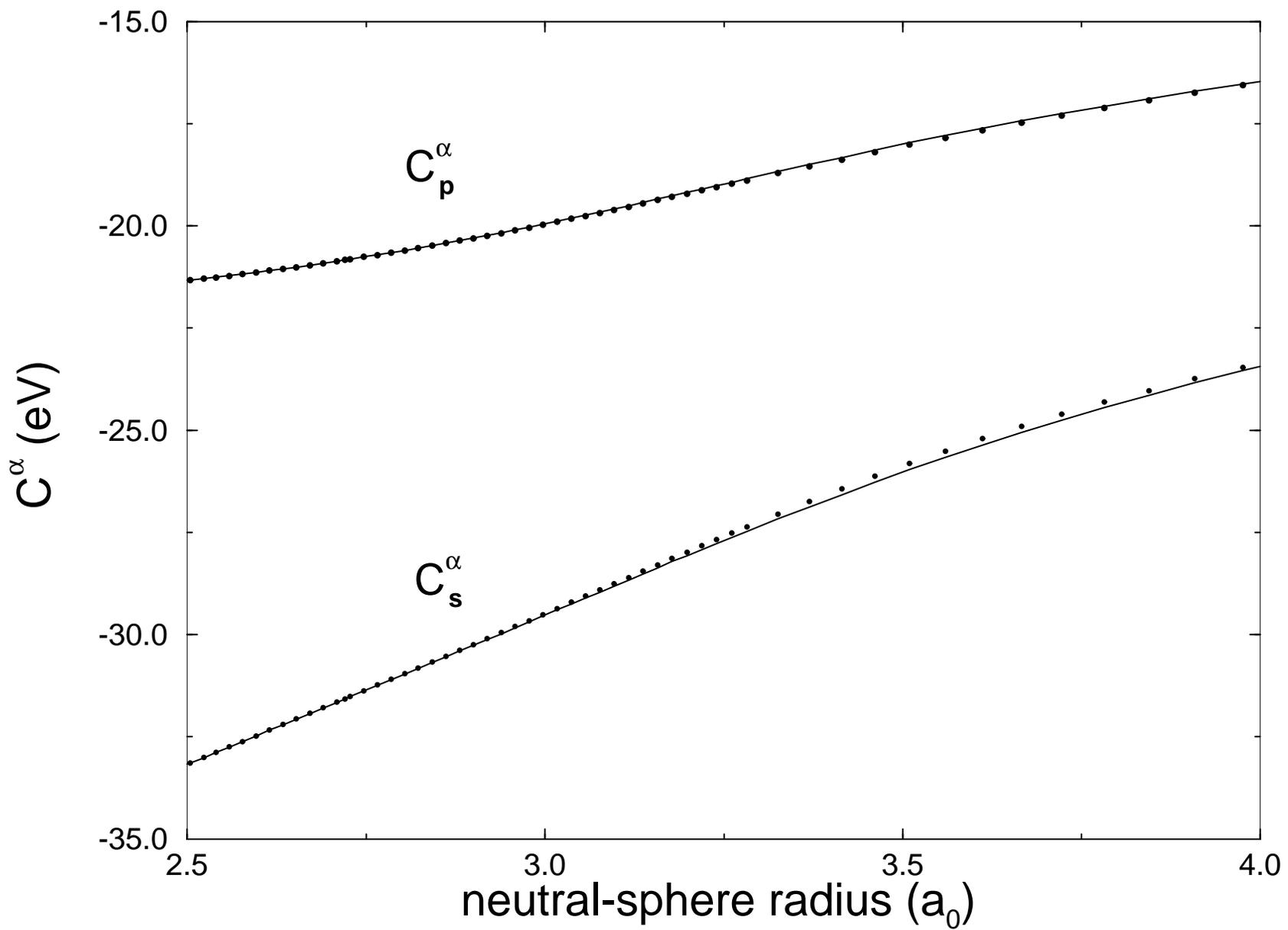
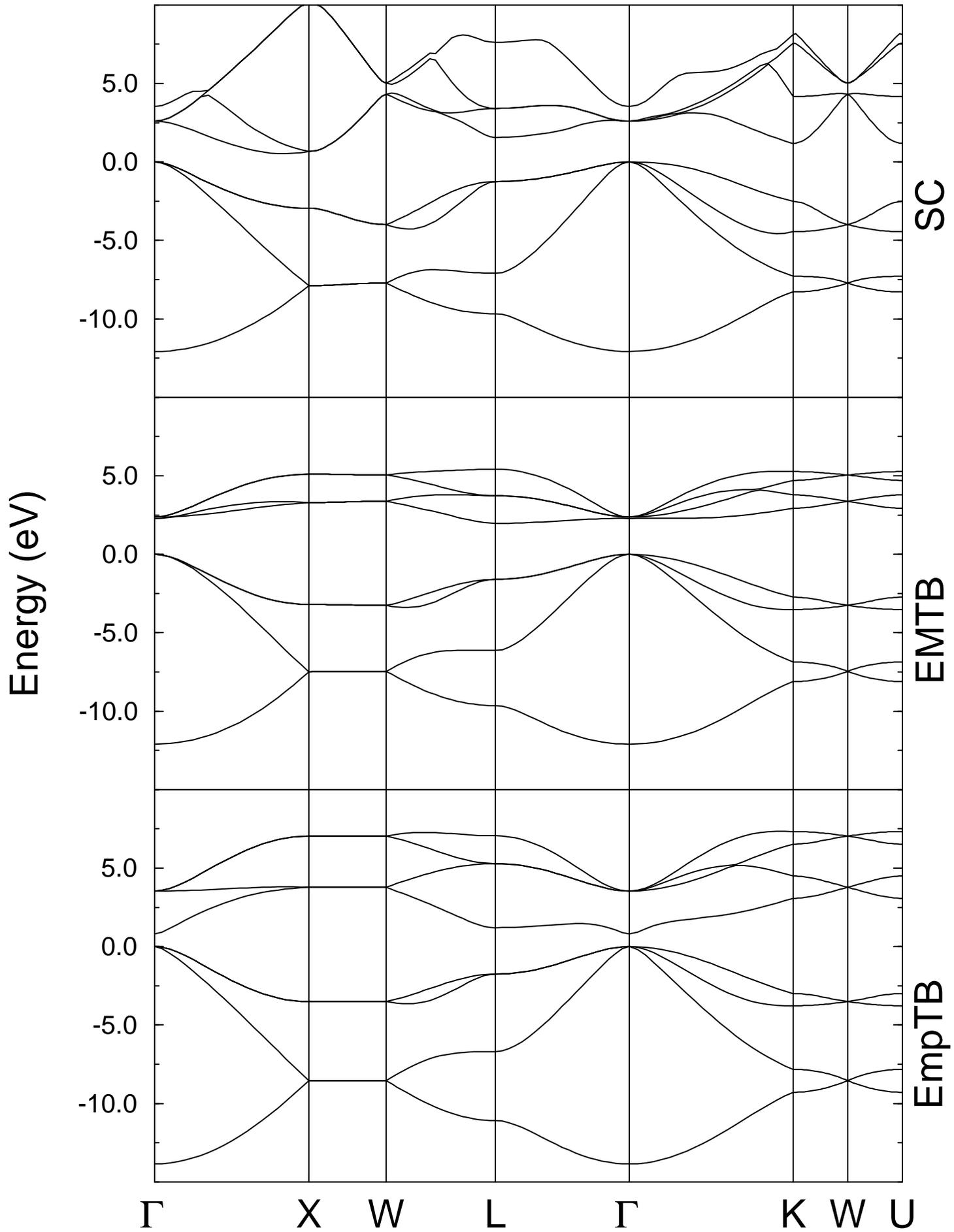
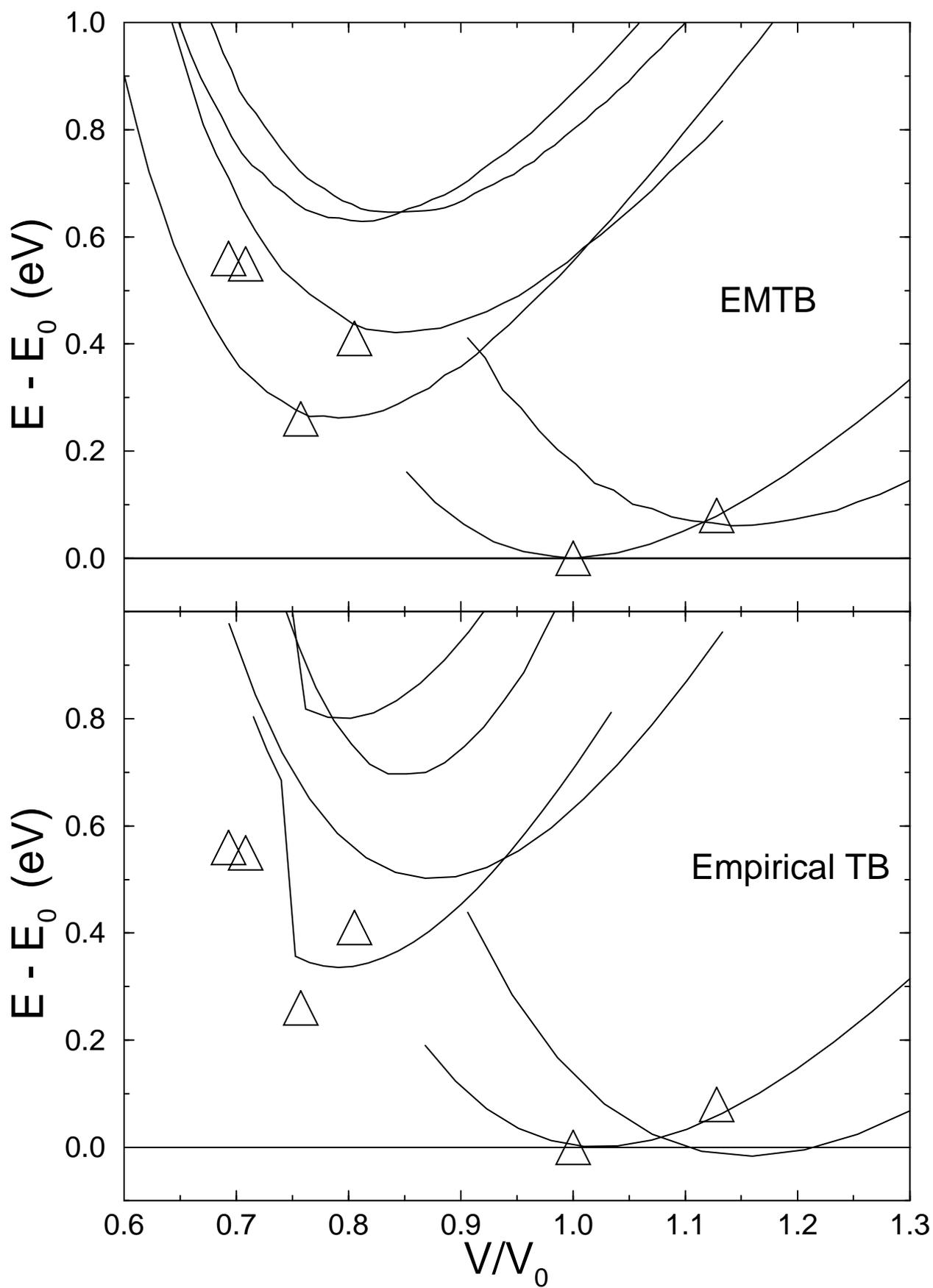
\end{document}